%% file: BDF.tex
\documentclass[a4paper,11pt]{article}
\pdfoutput=1 

\usepackage{jinstpub} 

\usepackage{color}   
\usepackage{hyperref}
\usepackage{ifthen}  
\newboolean{uprightparticles} 
\setboolean{uprightparticles}{false} 

\input{lhcb-symbols-def.tex}

\title{The~experimental~facility~for~the~Search~for~Hidden~Particles at the CERN SPS}

\emailAdd{Richard.Jacobsson@cern.ch}

\abstract{The Search for Hidden Particles (SHiP) Collaboration has shown that the CERN SPS accelerator with its $400 \gevc$ proton beam offers a unique opportunity to explore the Hidden Sector~\cite{ref:SHiP_TP,ref:SHiP_PP, ref:SHiP_TP_add}. The proposed experiment is an intensity frontier experiment which is capable of searching for hidden particles through both visible decays and through scattering signatures from
recoil of electrons or nuclei. The high-intensity experimental facility developed by the SHiP collaboration is based 
on a number of key features and developments which provide the possibility of probing a large part of the 
parameter space for a wide range of models with light long-lived super-weakly interacting particles with 
masses up to ${\cal O}(10) \gevcc$  in an environment of extremely clean background conditions. This paper describes the proposal for the experimental facility together with the most important feasibility studies. The paper 
focuses on the challenging new ideas behind the beam extraction and beam delivery, the proton beam dump, and the suppression of beam-induced background.}


\arxivnumber{1810.06880} 

\collaboration{\includegraphics[height=17mm]{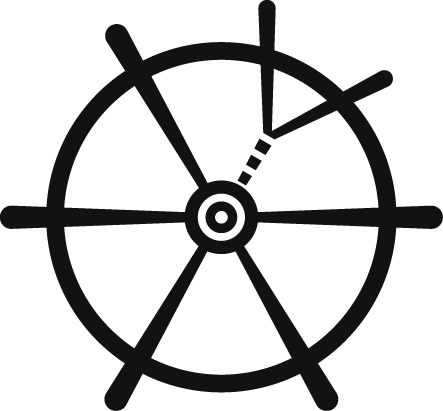}\\[6pt]
SHiP Collaboration\\
\vspace{2mm}
{\normalsize
C.~Ahdida$^{44}$,
R.~Albanese$^{14,a}$,
A.~Alexandrov$^{14}$,
A.~Anokhina$^{39}$,
S.~Aoki$^{18}$,
G.~Arduini$^{44}$,
E.~Atkin$^{38}$,
N.~Azorskiy$^{29}$,
J.J.~Back$^{54}$,
A.~Bagulya$^{32}$
% ...
\vspace{2mm}
}\newline
{\footnotesize \it
% ...
$ ^{14}$Sezione INFN di Napoli, Napoli, Italy\\
% ...
$ ^{18}$Kobe University, Kobe, Japan\\
% ...
$ ^{29}$Joint Institute for Nuclear Research (JINR), Dubna, Russia\\
% ...
$ ^{32}$P.N.~Lebedev Physical Institute (LPI), Moscow, Russia\\
% ...
$ ^{38}$National Research Nuclear University (MEPhI), Moscow, Russia\\
$ ^{39}$Skobeltsyn Institute of Nuclear Physics of Moscow State University (SINP MSU), Moscow, Russia\\
% ...
$ ^{44}$European Organization for Nuclear Research (CERN), Geneva, Switzerland\\
% ...
$ ^{54}$University of Warwick, Warwick, United Kingdom\\
% ...
$ ^{a}$Universit\`{a} di Bari, Bari, Italy\\
% ...
}}

\begin{document}
\maketitle
\flushbottom

\section{Introduction}
\label{sec:Introduction}

Given the absence of direct experimental evidence for Beyond the Standard Model (BSM) physics at the 
high-energy frontier and the lack of unambiguous experimental hints for the scale of new physics 
in precision measurements, it is plausible that the shortcomings of the Standard Model (SM) 
may have their origin in new physics only involving very weakly interacting, 
relatively light particles. Even in BSM scenarios associated with high mass scales such as in supersymmetry, many models contain light particles with suppressed couplings~\cite{ref:SUSY_light}. Considering the well-established observational evidence for a Hidden Sector 
in the form of Dark Matter, the structure and the phenomenology of the 
Hidden Sector may be more complex than just sourcing gravitational 
effects in the Universe. Non-minimal models of the Hidden Sector 
introduce various interactions and multiple types of hidden matter states  charged only under 
the hidden interactions, as well as various types of portal interactions between the visible sector 
of ordinary matter and the Hidden Sector
(\cite{ref:HS1}~-~\cite{ref:HS6}, \cite{ref:SHiP_PP} and references therein). 

As a consequence of the extremely feeble couplings for the portal interactions and the typically long lifetimes 
for the portal mediators, the low mass scales for hidden particles are far less constrained than the 
visible sector~\cite{ref:SHiP_PP,ref:HS6}. In several cases, the present experimental and theoretical constraints from cosmology 
and astrophysics indicate that a large fraction of the interesting parameter space was beyond the 
reach of previous searches, but is open and accessible to current and future facilities. While the 
mass range up to the kaon mass has been the subject of intensive searches, the bounds on the 
interaction strength of long-lived particles above this scale are significantly weaker.

Experimentally, the opportunity presents itself as an exploration at the intensity frontier with 
largest possible luminosity to overcome the very feeble interactions, and the largest possible acceptance 
to account for the typically long lifetimes. Beam-dump experiments are potentially superior to collider experiments in the sensitivity to GeV-scale hidden particles with their luminosities being 
several orders of magnitude larger than at colliders. The large forward boost for light states, giving good acceptance despite the smaller angular coverage and allowing efficient use of filters against background between the target and the detector, makes the beam-dump configuration ideal for searching for new particles with long lifetimes.

The recently proposed Search for Hidden Particles (SHiP) beam-dump experiment~\cite{ref:SHiP_TP} at the CERN Super Proton Synchrotron (SPS) 
accelerator is designed to both search for decay signatures by full reconstruction and particle 
identification of SM final states and to search for scattering signatures of Light Dark Matter by the detection of 
recoil of atomic electrons or nuclei in a heavy medium. Since the hidden particles, such as dark photons, dark scalars, heavy neutral leptons, and axion-like particles, are expected 
to be predominantly accessible through the decays of heavy hadrons and in radiative processes, 
the SHiP Collaboration has proposed an experimental facility which maximises their production and the detector 
acceptance while providing an extremely clean background environment. This paper focuses on describing the experimental facility.

 The proposal for the facility is based on a set of key themes.
Firstly, the full exploitation of the SPS accelerator with its present performance allows producing 
up to $2\cdot 10^{20}$ protons on target (Section~\ref{sec:pots}) in five years of 
nominal operation without affecting the operation of the Large Hadron Collider (LHC), and while maintaining the current level of beam usage for fixed-target facilities and test beam areas. 
The combination of the intensity and the $400\gev$ beam energy of the SPS 
proton beam produces yields of different light hidden sector particles which exceed those of existing or approved future facilities~\cite{ref:SHiP_TP_add}. 
At the same time, it has been found that the beam induced background flux at $400 \gev$ is manageable with the help of a hadron absorber and a muon shield system (Section~\ref{sec:muon_shield}). Secondly, the unique feature of slow extraction of a de-bunched beam over a timescale of around a 
second (Section~\ref{sec:slow_extraction}) allows a tight control of combinatorial background, and 
allows diluting the large beam power deposited on the proton target both spatially and temporally.

A set of innovative technological developments makes it possible to fully profit from these 
features. Several new techniques to improve the beam losses and irradiation inherent with slow beam 
extraction have been proposed and studied (Section~\ref{sec:slow_extraction}). Improvements in these areas are also 
of great interest to the existing CERN fixed target programs. The preliminary design of a long, complex, 
high-density primary proton target has been carried out (Section~\ref{sec:proton_target}). This target should be capable of 
coping with the large beam energy, and at the same time maximising the production of charm and beauty 
hadrons, and the production and interactions of photons, while minimising the production of 
neutrinos from pions and kaons. A yield of ${\cal O}(10^{18})$ charmed hadrons and ${\cal O}(10^{20})$ 
photons above $100~\mev$ are expected in five years of nominal operation. The feasibility of a target complex 
(Section~\ref{sec:target_complex}) which houses the proton target together with the associated services and remote 
handling, fully compatible with the radiation protection and environmental 
considerations, has been studied in detail. Furthermore, a new type of beam splitter magnet
(Section~\ref{sec:beamline}), which allows switching the beam to a short new transfer line to the SHiP 
experimental facility, while keeping all of the current experimental facilities in 
the CERN North Area operational, has been developed. The experimental configuration includes a unique design of a muon shield 
(Section~\ref{sec:muon_shield}) based on magnetic deflection to reduce the flux of muons by six 
orders of magnitude in the detector acceptance. A $\sim1700 \unit{m^3}$ experimental 
vacuum chamber (Section~\ref{sec:vac_vessel}), kept at a pressure of $1 \unit{mbar}$, allows suppressing residual 
neutrino-induced background.

Currently, CERN has no high-intensity experimental facility which is compatible with 
the full power of the SPS. CERN's North Area has a large space next to the SPS beam transfer lines 
which is for the most part free of structures and underground galleries, and which could accommodate the proposed 
facility. In addition, this facility is being designed with future extensions in mind.

At the energy of the SPS, the fully leptonic decays of the $D_s$ mesons are the principal 
source of tau neutrinos, with an expectation of ${\cal O}(10^{16})$ tau neutrinos in five years of nominal operation.
Thus, while the requirements for the experimental facility for the hidden particle search makes it unsuitable 
for neutrino oscillation physics, the setup allows studying interactions of tau and 
anti-tau neutrinos at unprecedented precision. With a ten-tonne $\nu$-target placed in front of the vacuum volume and equipped with suitable detectors, 
about $3\cdot 10^{4}$  $(2\cdot 10^{4})$ interactions of tau (anti-tau) neutrinos are expected 
within the geometrical acceptance. The first direct observation of 
the anti-tau neutrino and the measurement of tau neutrino and 
anti-tau neutrino cross-sections are 
among the physics goals of the proposed experiment. As charm hadron decays are also a source of 
electron and muon neutrinos, SHiP will also be able to study neutrino-induced charm 
production from all flavours with a dataset which is more than one order of magnitude larger than those collected by previous experiments. 

\section{Experimental set-up}
\label{sec:exp_setup}

\begin{figure}[tbh]
\centering
\includegraphics[width=0.95\linewidth]{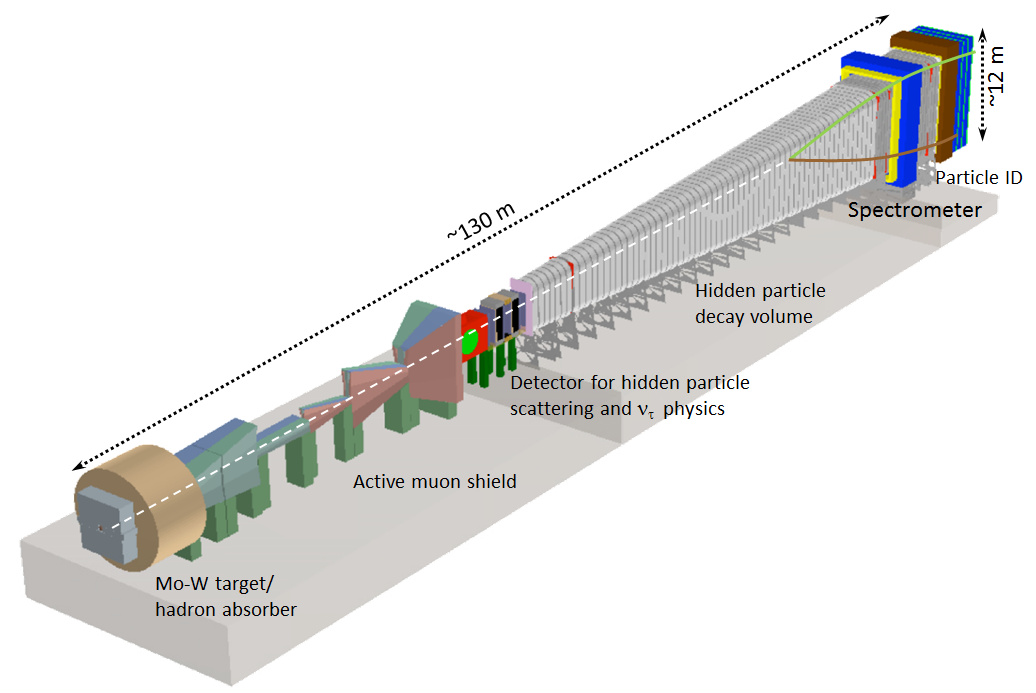}
\caption{Overview of the target and experimental area for the SHiP detector as implemented in the physics simulation.}
\label{fig:SHiP_facility_overview}
\end{figure}

\begin{figure}[tbh]
\centering
\includegraphics[width=0.99\linewidth]{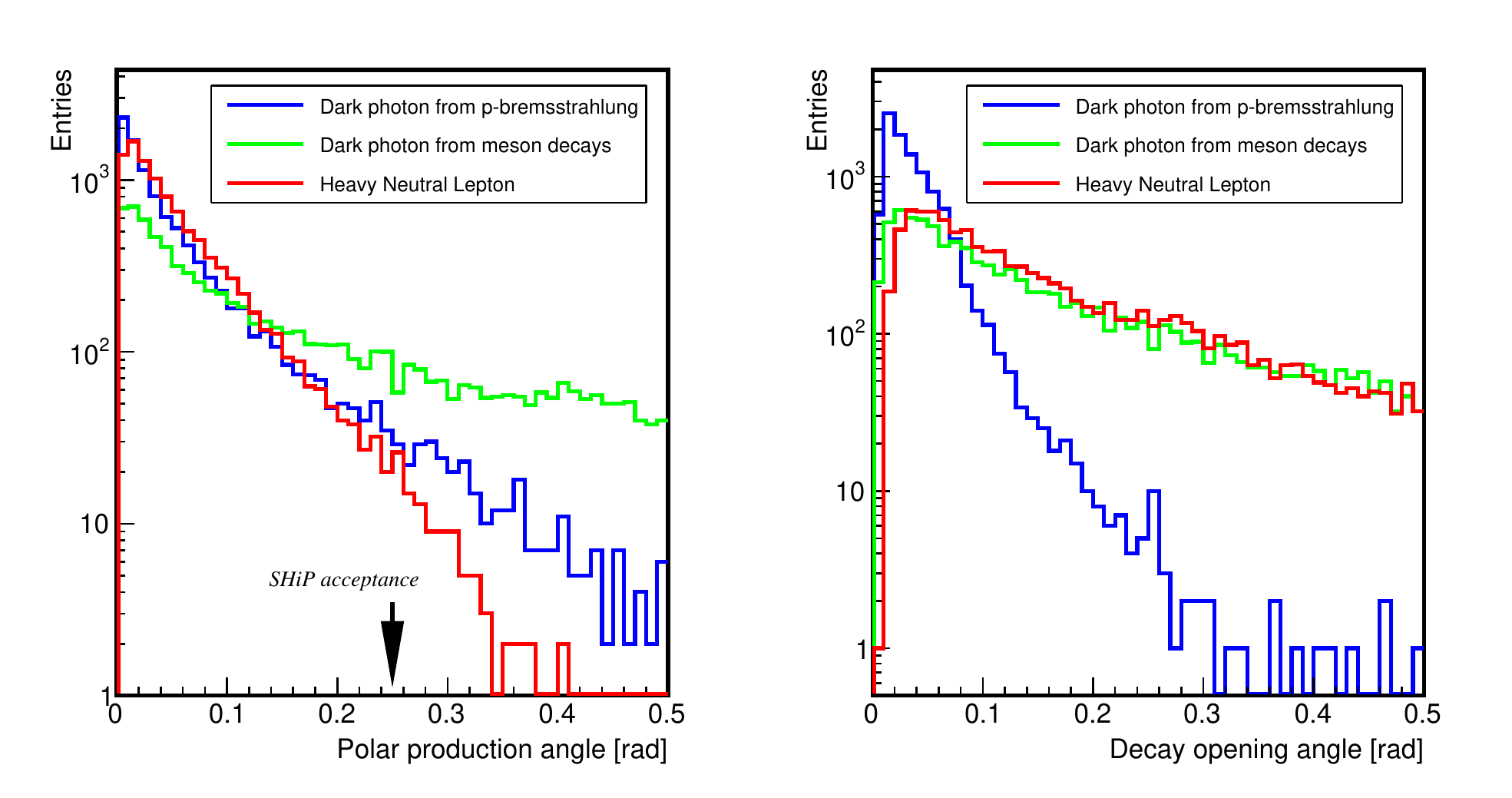}
\caption{(Left) Polar production angle with a beam momentum of $400\gevc$ for dark photons ($A$) produced in proton bremsstrahlung ($m_A = 2.0\gevcc$) and in meson decays ($m_A = 0.9\gevcc$),  and for heavy neutral leptons (HNL) ($m_{HNL} = 1.0\gevcc$) from decays of charm hadrons. The arrow indicates the acceptance of the SHIP fiducial volume, given by the transverse size of the decay volume (Right) Decay opening angles for two-body decays of the same three cases. The geometry of the decay volume has been optimized given the aperture of the spectrometer and the hidden particle kinematics.}
\label{fig:hs_kinematics}
\end{figure}

The experimental requirements, as dictated by the phenomenologies of the different Hidden Sector models, are very similar. This allows the design of a general-purpose layout based on a global optimisation of the experimental facility and of the SHiP detector.
Figure~\ref{fig:SHiP_facility_overview} 
shows an overview of the experimental facility from the proton target to the end of the Hidden Sector detector. 
The main challenges concern the requirement of a highly efficient reduction of the very large beam-induced background, and an efficient and redundant 
tagging of the residual background down to below $0.1$ events in the projected sample of 
$2\cdot 10^{20}$ protons on target. Despite the aim to cover long lifetimes, the sensitive volume should be situated as close as possible to the proton target due to the relatively large transverse
momentum of the hidden particles resulting from the limited boost of the heavy hadrons 
(Figure~\ref{fig:hs_kinematics}). The minimum distance is only constrained by the need 
of a system to absorb the electromagnetic radiation and hadrons emerging from the proton target and to reduce the beam-induced muon flux. 

The proton target, described in Section~\ref{sec:proton_target}, is followed by a $5\unit{m}$ long hadron absorber. The physical 
dimensions of the absorber are mainly driven by the radiological requirements. In addition to absorbing the hadrons 
and the electromagnetic radiation, the iron of the hadron absorber is magnetised over a length of $4\unit{m}$. The applied dipole field makes up the first section of the active muon shield (Section~\ref{sec:muon_shield}) which is optimised to sweep out of acceptance the entire spectrum of muons up to $350\gevc$. 
The remaining part of the muon shield follows immediately downstream of the hadron absorber in the experimental hall and consists of a chain of magnets which extends over a length of $\sim 40\unit{m}$.

The SHiP experiment incorporates two complementary apparatuses. The detector system immediately downstream of the muon shield is optimised both for recoil signatures of hidden sector particle scattering and for neutrino physics. It is based on a hybrid detector similar to what was developed by the OPERA Collaboration~\cite{ref:opera} with alternating layers of nuclear emulsion films and electronic trackers, and high-density $\nu$-target plates. In addition, the detector is located in a magnetic field for charge and momentum measurement of hadronic final states. The detector 
$\nu$-target mass totals ${\cal O}(10)\unit{tonnes}$. The emulsion spectrometer is followed by a muon 
identification system. This also acts as a tagger for interactions in the muon filters which may 
produce long-lived neutral mesons entering the downstream decay volume and whose decay may mimic 
signal events. 

The second detector system aims at measuring the visible decays of Hidden Sector particles to both 
fully reconstructible final states and to partially reconstructible final states with neutrinos.
The detector consists of a $50\m$ long decay volume (Section \ref{sec:vac_vessel}) followed by a 
large spectrometer with a rectangular acceptance of $5\m$ in width and $10\m$ in height. The length of the decay volume is 
defined by maximising the acceptance to the hidden particle decay products (Figure~\ref{fig:hs_kinematics}) given the transverse 
size of the spectrometer. In order to suppress the background from 
neutrinos interacting in the fiducial volume, it is maintained at a pressure of 
${\cal O}(10^{-3})\unit{bar}$. The spectrometer is designed to accurately reconstruct the decay vertex, 
the mass, and the impact parameter of the hidden particle trajectory at the proton target. A set of 
calorimeters and muon stations provide particle identification. The system is optimised to detect as 
many final states as possible in order to be sensitive to, and discriminate between, a very 
wide range of models. A dedicated timing detector with $\sim100\ps$ resolution provides a 
measure of coincidence in order to reject combinatorial backgrounds. The decay volume is 
surrounded by background taggers to identify neutrino and muon inelastic scattering in the vacuum 
vessel walls which may produce long-lived neutral SM particles, such as $K_L$ etc. 

The muon shield and the SHiP detector systems are housed in a $\sim120\m$ long underground 
experimental hall at a depth of $\sim15\m$. To minimise the background induced by the flux of 
muons and neutrinos interacting with material in the vicinity of the detector, no infrastructure
systems are located on the sides of the detector, and the hall is $20\m$ wide along the 
entire length. 

\begin{figure}[tbh]
\centering
\includegraphics[width=0.87\linewidth]{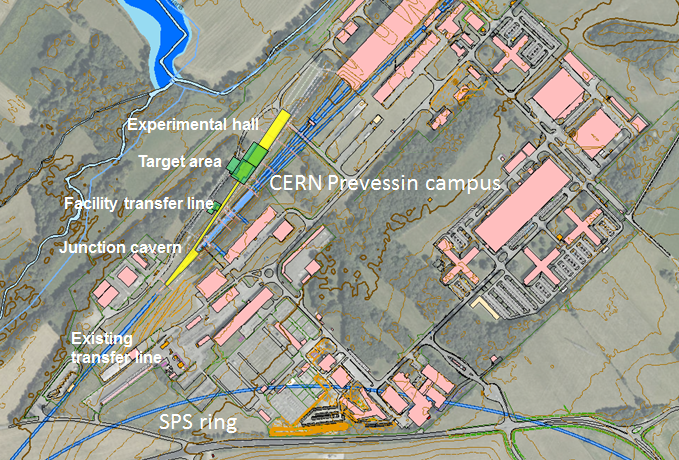}
\caption{Overview of the required civil engineering for the proposed experimental facility for SHiP on the CERN Prevessin campus. 
The beam-axis is at a depth of about $10\m$ which allows trenching the entire complex from the surface.
New or reworked construction in yellow (underground) and green (surface); existing tunnels in blue.}
\label{fig:facility_ce_works}
\end{figure}

Figure~\ref{fig:facility_ce_works} shows an overview of the civil engineering required for the 
experimental facility for SHiP. All civil engineering works are fully located within existing CERN 
land on the Prevessin campus. This location is very well suited to house the experimental
facility, owing to the stable and well understood ground conditions, accessible services 
and very limited interference with existing buildings, galleries and road structures. By maintaining the entire beam line horizontal and at the same level as the existing splitter 
region at the end of the SPS extraction line, the experimental hall is conveniently situated 
at a depth of about $15\m$, which is compatible with the requirements from radiation protection while still 
allowing easy direct access from above without a shaft. 

\section{Proton beam}
\label{sec:proton_beam}
\begin{figure}[tbh]
\centering
\includegraphics[width=0.8\linewidth]{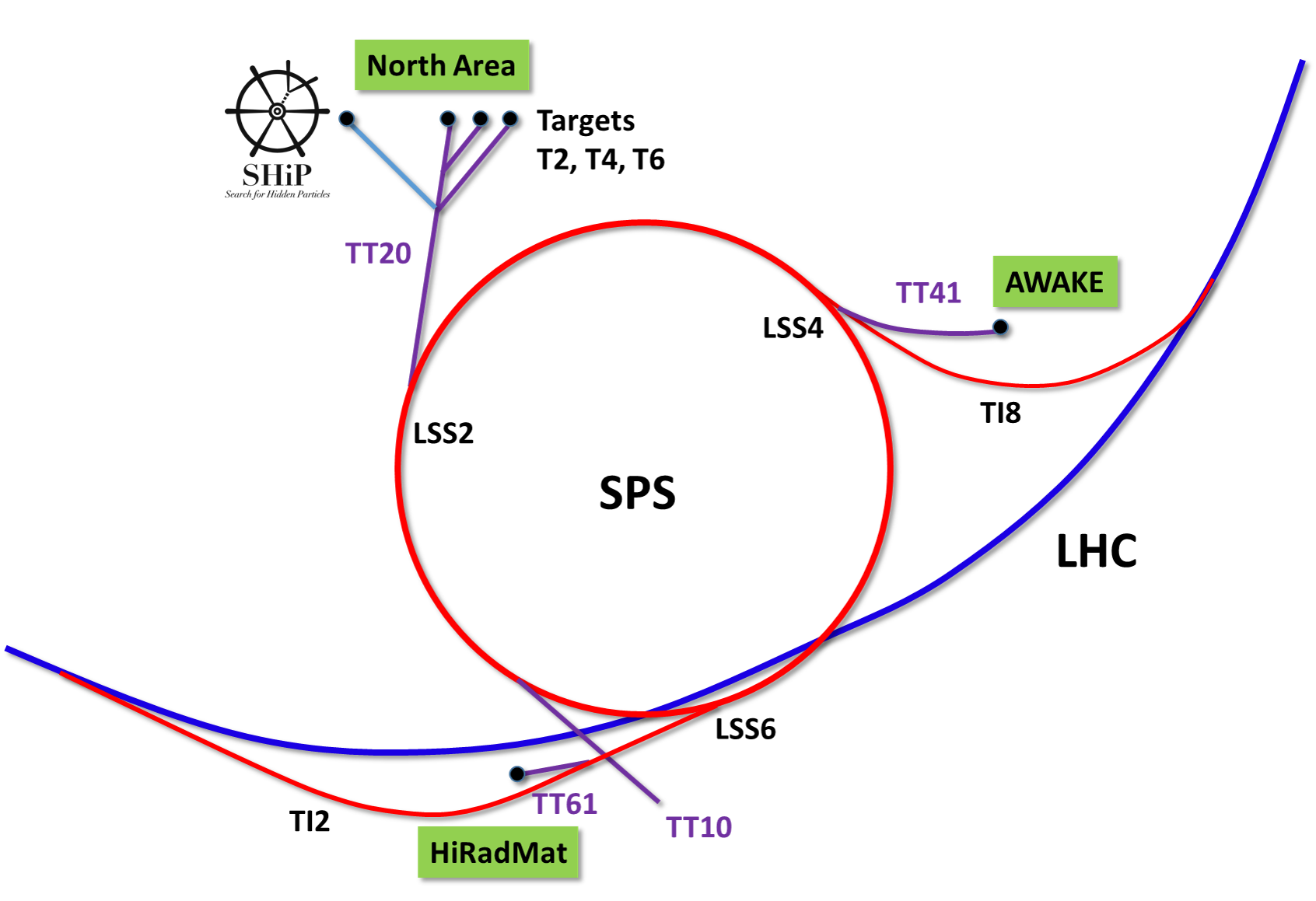}
\caption{Overview of the SPS accelerator complex. The SHiP experimental facility is located
in the North Area and shares the TT20 transfer line with the fixed target programs.}
\label{fig:facility_acc_complex}
\end{figure}

The proposed implementation of the SHiP experimental facility is based on minimal modifications to the SPS 
complex and a maximum use of the existing accelerator and beam lines. Figure~\ref{fig:facility_acc_complex} 
shows schematically the proposed location of the experimental facility at the CERN North Area site. 
The facility shares about $600\m$ of the existing TT20 transfer line with the other North Area 
facilities. 

At the SPS, the most favourable experimental conditions for SHiP are obtained with a proton beam momentum of around 
$400\gevc$. Based on the SPS in its current state and in view of its past performance, a nominal beam intensity
of $4\cdot 10^{13}$ protons on target per spill is assumed for the design of the experimental facility 
and the detector. 

\begin{figure}[tbh]
\centering
\includegraphics[width=0.6\linewidth]{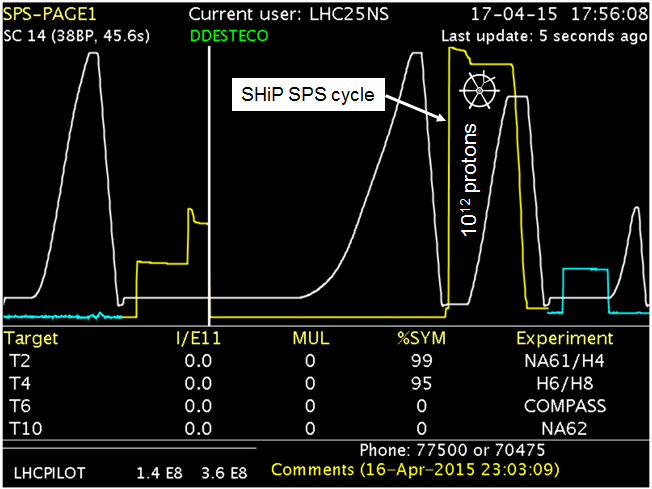}
\caption{First slow beam extraction tests from the SPS for SHiP with the specific length of about $1\sec$. The tests were performed at low intensity of about $10^{12}$ protons/s. The yellow line represents the proton beam intensity in the SPS and the white line represents the SPS beam energy.}
\label{fig:SHiP_cycle}
\end{figure}

In order to reduce the probability of combinatorial background events from 
residual muons entering the detector decay volume and to respect the limits on the instantaneous beam
power deposited in the proton target, SHiP takes advantage of the SPS slow extraction used to 
provide beam to the CERN North Area through the Long Straight Section 2 of the SPS. The minimum 
SPS cycle length which is compatible with these requirements is $7.2\sec$. A beam cycle with a slow 
extraction of around one second has already been demonstrated in the studies for the experimental facility for SHiP (Figure~\ref{fig:SHiP_cycle}).

\subsection{Achievable protons on target and beam sharing}
\label{sec:pots}

\begin{figure}[tbh]
\centering
\includegraphics[width=0.8\linewidth]{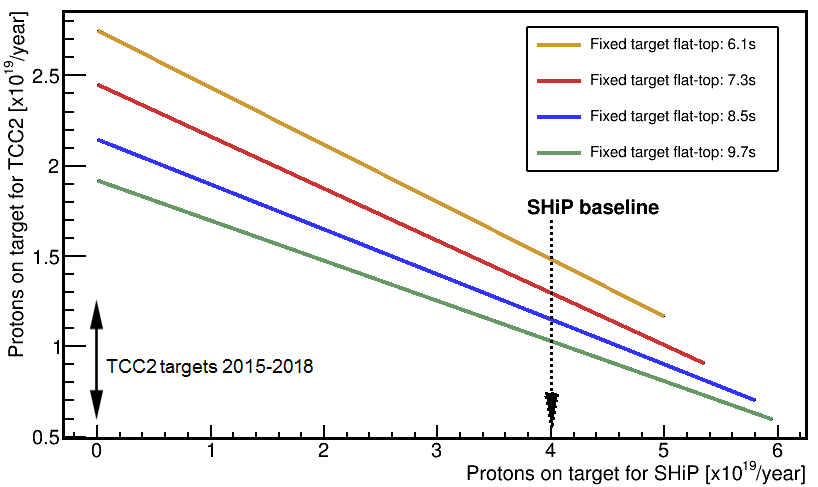}
\caption{The expected number of protons on the current North Area targets (TCC2) as a function of the 
number of protons on target for the SHiP experimental facility with a 1.2\,s spill length. 
The plot shows the performance for different 
spill durations for the current fixed target facilities between 6.1 -- 9.7s. The range of the numbers of
protons per year delivered to the North Area targets in the years 2015 -- 2018 is indicated. The preferred working point for SHIP is indicated by "SHiP baseline".}
\label{fig:facility_pot}
\end{figure}

The SHiP operational scenario is based on a similar fraction of beam time as the recently 
completed CERN Neutrinos to Gran Sasso (CNGS) program, and assumes the operational performance 
of the SPS in recent years~\cite{ref:CNGS_performance}. Compatibility with the existing North Area program is important, and Figure~\ref{fig:facility_pot} shows the number of protons on the 
current North Area targets as a function of the number of protons on the SHiP proton target for 217 
days of physics, corresponding to the situation for the 2011 run. It has been assumed that 
10\% of the SPS scheduled physics time is devoted to run LHC pilot cycles and another 
10\% to run LHC nominal cycles. The assumed sharing delivers an annual yield of 
$4\cdot 10^{19}$ protons on target to the SHiP experimental facility and a total of $1\cdot 10^{19}$ 
to the other physics programs at the CERN North Area. The physics sensitivities of the 
experiment are calculated based on acquiring a total of $2\cdot 10^{20}$ protons on target 
which may thus be achieved in five years of nominal operation.

\subsection{Extraction beam loss and activation}
\label{sec:slow_extraction}

\begin{figure}[tbh]
\centering
\includegraphics[width=0.7\linewidth]{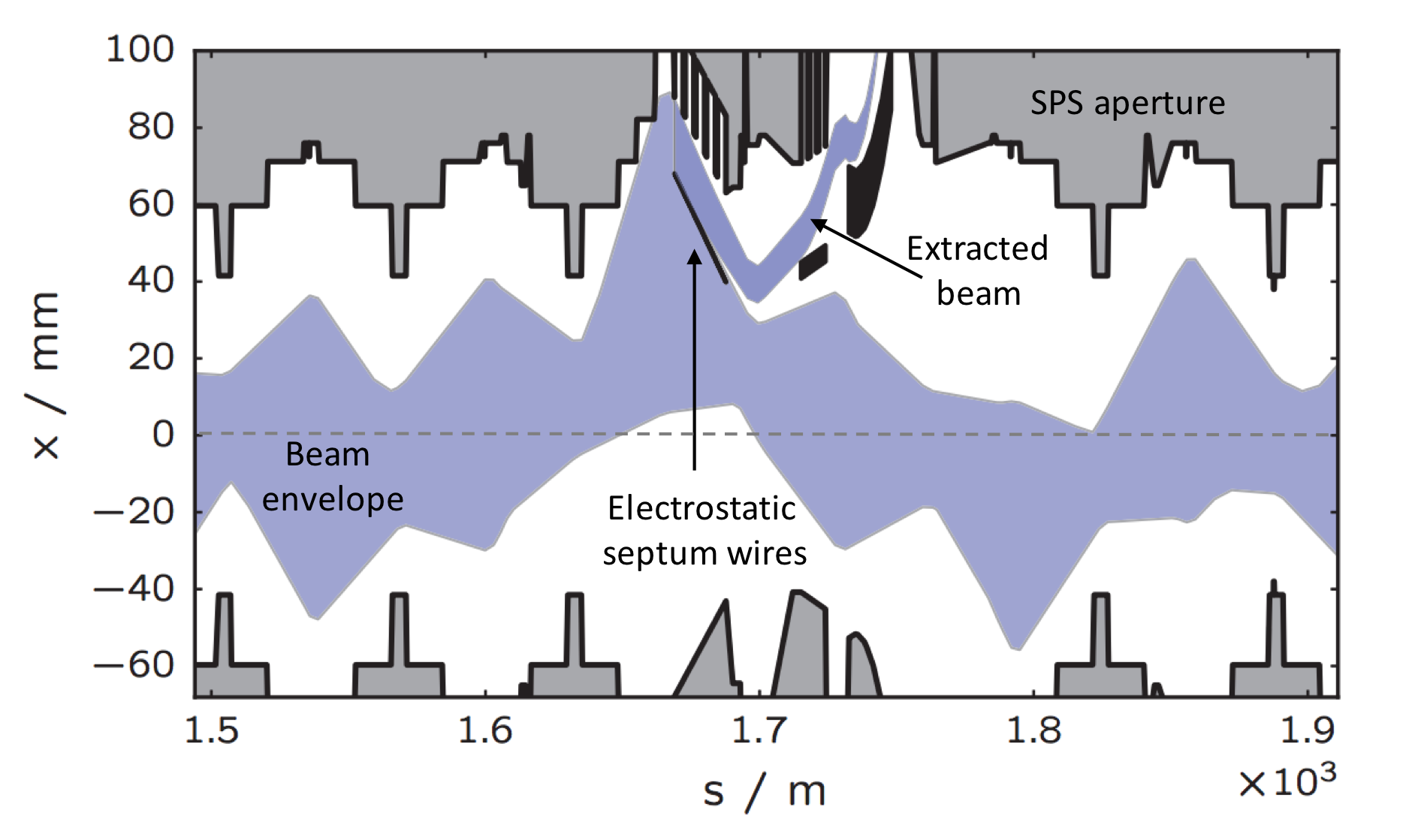}
\caption{Envelope of the circulating and the extracted beam along the SPS extraction region, showing the machine aperture and the wires of the electrostatic septum. The passive diffuser or bent crystal are located just upstream of the electrostatic septum to reduce the density of protons impacting the wire.}
\label{fig:extraction}
\end{figure}

The slow extraction from the SPS exploits a third-order resonance 
to achieve a controlled continuous amplitude growth of the transverse oscillations of 
the circulating protons. The amplitudes grow over several tens of thousands of turns 
until a slice of the beam
crosses the wires of the electrostatic septa, and is guided into the 
TT20 beamline aperture continuously, as shown in Figure~\ref{fig:extraction}, until 
the circulating beam in the SPS is completely extracted.  
The field wires have finite width and inevitably intercept a fraction of the beam, 
leading to beam losses of the order of 2\% of the total intensity. This is an important difference with 
respect to CNGS operation, which used essentially loss-free fast extraction. 

In addition to the increased risk of sparking and damage to the wires due to heating and vacuum pressure rise, the main consequences of beam loss are radio-activation of 
the extraction region, accumulated radiation damage to sensitive equipment 
and cables, and the increased cool-down times in case of interventions for repair or maintenance. Activation and personnel dose is already a serious issue in the SPS, and currently reach operational limits with around $1.2\cdot10^{19}$ protons slowly extracted per year. 

To extrapolate to the operation of the experimental facility for SHiP, the experience from operating the West Area Neutrino Facility (WANF) has been studied.
Approximately half of the total integrated number of protons foreseen for SHiP was extracted to WANF with fast-slow (half-integer) extraction during a five-year period at the end of the 1990's.
More recent experience of sending beam to the North Area has also been considered, e.g. over $2\cdot10^{19}$ protons were slowly extracted to the North Area during 2007.
The studies show that a factor of four decrease in the potential radiation dose to personnel is required to achieve the SHiP baseline intensity of $4\cdot10^{19}$ protons on target per year. 
This improvement will need to come from a combination of reduced beam loss, reduced activation per lost proton, and improved or remote interventions.

Extraction losses have been improved already by 
increasing the stability of the extraction with the help of a feed-forward system on the main quadrupole current 
to compensate for the ripple induced by the main electricity grid. Also, the septum wires are regularly realigned with 
the help of improved instrumentation and algorithms. However, a significant decrease (i.e. a factor two or more) can only be expected with 
substantial changes to the extraction dynamics. Studies involving two techniques based on coherent and incoherent 
scattering of the protons upstream of the septum (Figure~\ref{fig:extraction}), that would otherwise hit the septum wires, are currently being tested, along with ways of modifying 
the transverse phase space distribution to reduce particle density at the wires. 

\begin{figure}[tbh]
\centering
\includegraphics[width=0.6\linewidth]{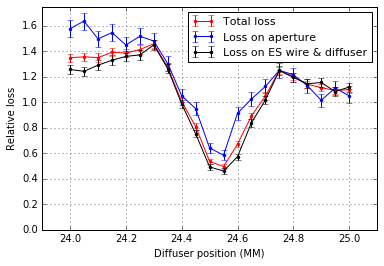}
\caption{Relative loss of protons in arbitrary units as a function of the transverse position from simulation of a $3\mm$ long, $0.24\mm$ wide tungsten-rhenium diffuser. The sum of the loss on the diffuser and the electro-static septum (ES) wires is lower than the total loss with ES wires alone, because the scattering from the diffuser reduces the particle density at the ES sufficiently to result in an overall loss reduction. A factor two improvement is obtained for the optimal position.}
\label{fig:diffuser}
\end{figure}

The first technique is based on a passive beam scattering device. It consists of a short, thin blade of a
high-Z material located upstream of the electrostatic septum wires. The blade intercepts a thin 
slice of the beam in order to generate an angular spread which reduces the transverse beam 
density at the wires, resulting in an overall reduction of the beam losses. Simulations show that this technique could 
bring up to a factor two improvement (Figure~\ref{fig:diffuser}). The device is also straightforward to deploy and 
operate. A prototype diffuser to benchmark the simulations with
experiment is being designed and built. It will be installed in the SPS Long Straight Section 2 (LSS2 in Figure~\ref{fig:facility_acc_complex}) and tested with beam in 2018.

The second technique employs a thin bent crystal placed upstream of the septum in order to 
channel away the misdirected protons into the extraction aperture. Since the
channeling is very sensitive to the angular alignment of the crystal, the efficiency of 
this technique depends strongly on the angular spread of the beam 
and the orbit stability. A proof-of-principle experiment with coasting beam has already demonstrated~\cite{ref:crystal_extraction}
that beam can be extracted into the TT20 transfer line using a bent crystal. 

Both the crystal-assisted slow extraction and the diffuser rely on stable 
conditions and an accurate alignment of the septum wires and the scattering device. A movement of the 
extraction separatrix in position and more importantly angle is, however, inherent to the SPS extraction
mechanism optimised for low beam loss. Use of a dynamic extraction bump could compensate in real-time for these changes in the closed orbit. This could also permit a faster realignment of the beam with the septa, instead of the time-consuming mechanical realignment of the septa.

A final set of studies focuses on manipulation of the transverse phase-space distribution, using either higher-order
multipole magnets or a pair of septum elements in which the configuration of the conductor and magnetic material is used to separate the high-field region from the zero-field region without intervening physical material ("massless septa"), to reduce the particle density at the septum wires without increasing 
losses elsewhere in the extraction system. These approaches are being studied in simulation and proof-of-principle 
measurements have been planned for 2018. First studies of combining these techniques with the diffuser, or the crystal, indicate that it can potentially improve the loss reduction well beyond a factor two.

The different mitigation techniques are also complemented by studies of alternative materials for construction of septum 
sub-systems like titanium or carbon nanotubes to reduce activation, and developments of machine assisted intervention techniques. 

\subsection{Spill harmonic content}

Suppression of combinatorial background from residual muons produced in the SHiP proton target rely on determining the time coincidence of the reconstructed tracks in the SHiP spectrometer with the help of a timing detector. The requirement on time resolution is derived from the likelihood of coincidental muons. The likelihood is directly related to the proton interaction rate in the target, which should have minimal variations. The baseline beam parameters and the average residual muon flux in 
the detector acceptance requires a timing detector with a time resolution of ${\cal O}(100)\ps$. Rejection of combinatorial background is thus one of the main drivers for a highly uniform extraction of the spill.

In 2017 sample spills were generated with the SHiP beam cycle, with the encouraging result that the spill harmonic content is not worse than for the 
longer spills used for the North Area. Contributions are dominated at low frequency by the effect of harmonics on the main electricity grid affecting the extraction beam dynamics. 
To this end, improvements of the stability of the slow 
extraction are also aiming at improving the uniformity of the spill structure.
At higher frequencies the residual radio frequency structure of the beam dominates. 

\subsection{Beamline to proton target}
\label{sec:beamline}

\begin{figure}[tbh]
\centering
\includegraphics[width=0.8\linewidth]{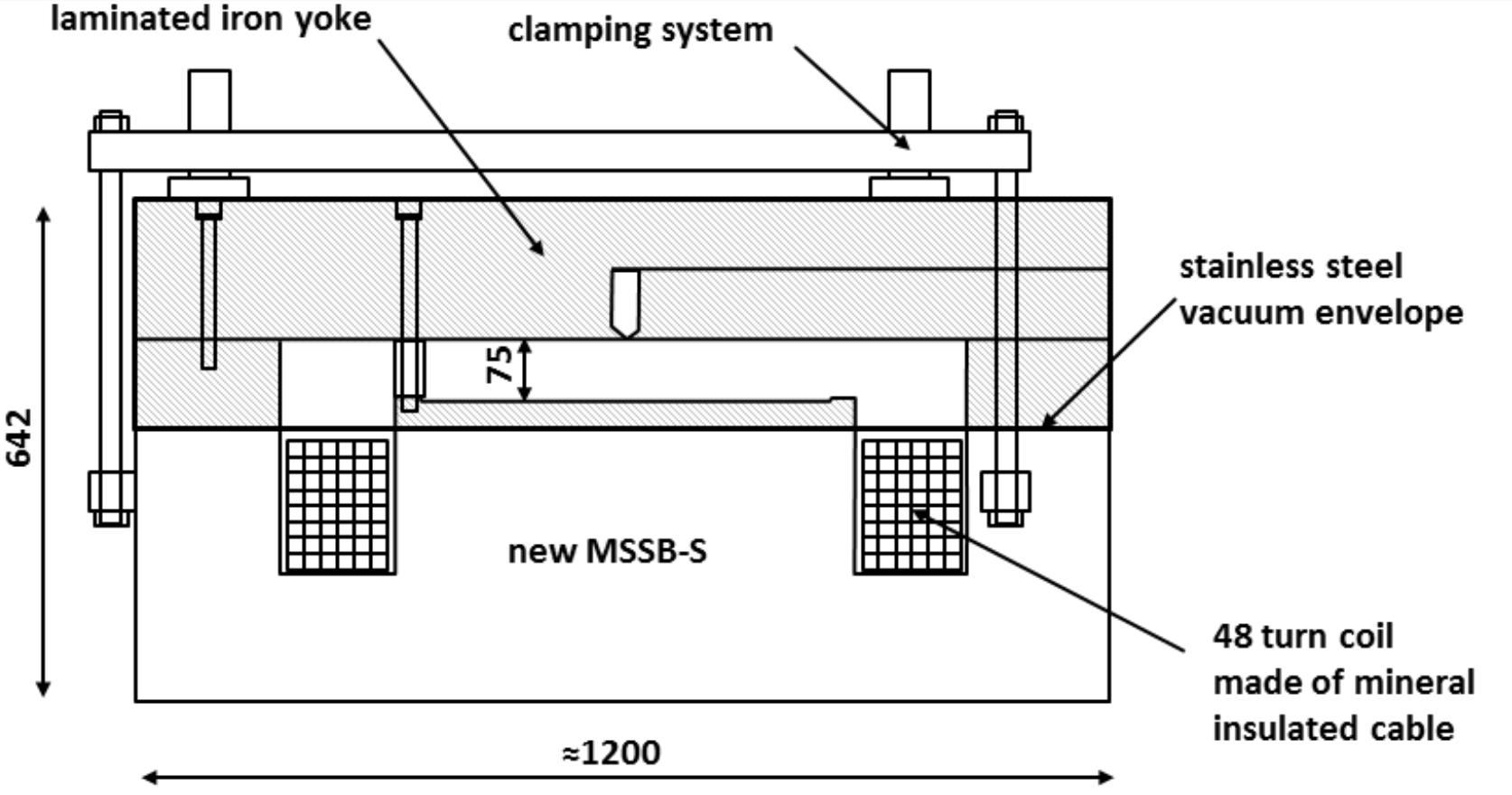}
\caption{Cross-section of the new "MSSB-S" splitter magnet. The cycle-to-cycle polarity switching requires a laminated iron yoke. The $7.5\mm$ beam gap is made significantly wider than in the original splitter and extends to both sides of the septum to accommodate both the deflection of the SHiP beam to one side and alternatively splitting the beam between the other North Area facilities on the other side. All dimensions are in $\mm$.}
\label{fig:splitter}
\end{figure}

The location of the SHiP proton target in the North Area allows the re-use of about $600\m$ of the present TT20 transfer line, which has sufficient aperture for the slow-extracted beam at $400\gevc$. The new dedicated beam transfer line to the experimental facility for SHiP branches off at the end of the TT20 transfer line with the help of a set of newly proposed bi-polar splitter magnets which replaces the existing ones. The new magnets allow both maintaining the present function of splitting the 
beam between the proton target for the experimental area currently hosting the COMPASS 
experiment~\cite{ref:COMPASS_experiment} and the rest of the existing North Area facilities, and to alternatively 
switch the entire spill to the dedicated transfer line for SHiP, on a cycle-by-cycle basis. The present magnet is an in-vacuum Lambertson septum with a yoke machined from solid iron, with the coil based on a water-cooled lead of copper with an insulation of compacted MgO powder~\cite{ref:splitter_magnets}. For the new magnets a laminated yoke is required in order to rapidly perform the polarity switch between SPS cycles, which implies ramping the field reliably in about $2\sec$. The new magnets, shown in Figure~\ref{fig:splitter}, must also have a larger horizontal aperture, as the beam is deflected to different sides of the magnet axis for SHiP and for North Area operation. R\&D and prototyping of the laminated yoke is underway to study the very tight mechanical 
tolerances required in the septum region in order to maintain low beam losses. Similar MgO coil technology as used in the existing splitter will provide the required radiation resistance.

\begin{figure}[t]
\centering
\includegraphics[width=0.6\linewidth]{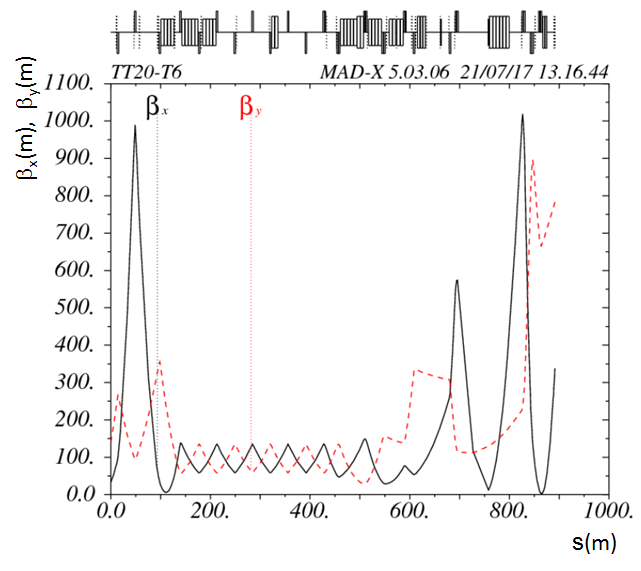}
\caption{Optics shown by the beta function in the horizontal (black solid line) and vertical plane (red dashed line) along the entire length of the  beamline from the SPS extraction ($s=0\m$) in LSS2 to the SHiP proton target located at around $s=900\m$. The new section of beamline is matched to the existing TT20 line to give the required beam size at the target.}
\label{fig:optics}
\end{figure}

A $380\m$ long new section of beam line is needed, which is matched to the existing TT20 transfer line (Figure~\ref{fig:optics}) and which brings the beam up to the new target complex. A preliminary design has been made which exploits 17 standard SPS warm bending magnets, running at a conservative field of $1.73\unit{T}$ producing an angular beam deflection of $8\unit{mrad}$ each, to increase as much as possible the distance between the new and existing beam lines. A maximum deflection angle to exit from the tunnel of the existing beam line is beneficial to reduce the
longitudinal extent of the civil engineering works in the crucial junction region. The bending dipoles
downstream of the splitter are grouped into a single dipole unit as early as possible, with four subsequent
standard SPS half-cells of four dipoles, each separated by a quadrupole. The
powering scheme for the TT20 transfer line remains largely unchanged up to the switch element
with cycle-to-cycle rematching of the last nine quadrupoles before the splitter and steering,
to allow the entire beam cross-section to pass through the dipole aperture with very low
losses. The quadrupoles in TT20 are
already laminated and suited to cycle-to-cycle switching.  

For the new beam line, around six new corrector dipoles are assumed. In addition, five standard SPS quadrupole magnets will be required to control the vertical beam size
through the dipole apertures, and provide flexibility and tunability of the beam spot size
and dispersion at the proton target. In order to produce sufficient dilution of the beam power in the SHiP proton target, the slow extraction is combined with a beam spot of at least $6\mm$ root-mean square in both planes and a large sweep 
of the beam over the target surface. The beam sweep is implemented with two orthogonal 
kicker magnets located after the last bending dipole 
magnet at $120\m$ upstream of the target, with Lissajous powering functions to produce a circular sweep. With a 
free drift length for the beam of about $120\m$ and a bending angle of $0.25\unit{mrad}$ per plane, it is possible to achieve a sweep radius of $30\mm$. Since the survival of the proton target relies critically on the beam 
dilution, the SPS beam is interlocked with the beam dilution system and the instantaneous loss rate at the target.

The overall layout and clearances allow civil engineering to take place along the 
entire experimental facility starting from the middle of the new transfer line and up to the end of the experimental hall during beam 
operation for the other North Area facilities.

\section{Proton target and target complex}
\label{sec:beam_dump}

\subsection{Design constraints for the proton target}
\label{sec:proton_target}

The physics scope of the SHiP experiment requires a proton target which maximises the production 
of photons, and $D$ and $B$ mesons. At the same time, the proton interactions give 
rise to copious direct production of short-lived meson resonances, as well as pions and kaons. While 
a hadron absorber of a few meters of iron is sufficient to absorb the hadrons and the 
electromagnetic radiation emerging from the target, the decays of the pions, kaons and 
short-lived meson resonances result in a large flux of muons and neutrinos. In order to reduce the 
flux of neutrinos, in particular the flux of muon neutrinos and the associated muons,
the pions and kaons should be stopped as efficiently as possible before they decay.  The target 
should thus be made of a material with the highest possible atomic mass and atomic charge. It should be sufficiently long to intercept virtually all of the
proton intensity and to contain the majority of the hadronic shower with minimum leakage. 
Simulation~\cite{ref:cascade} shows that re-interactions of primary protons and interactions 
of secondaries produced in the nuclear cascades contribute with a significant amplification of 
the signal yields. For instance, in the case of charm production, the cascade processes 
contribute by more than doubling the yield as compared to what is expected from only the 
primary proton-nucleus interactions.

\begin{figure}[tbh]
\centering 
\includegraphics[width=0.85\linewidth]{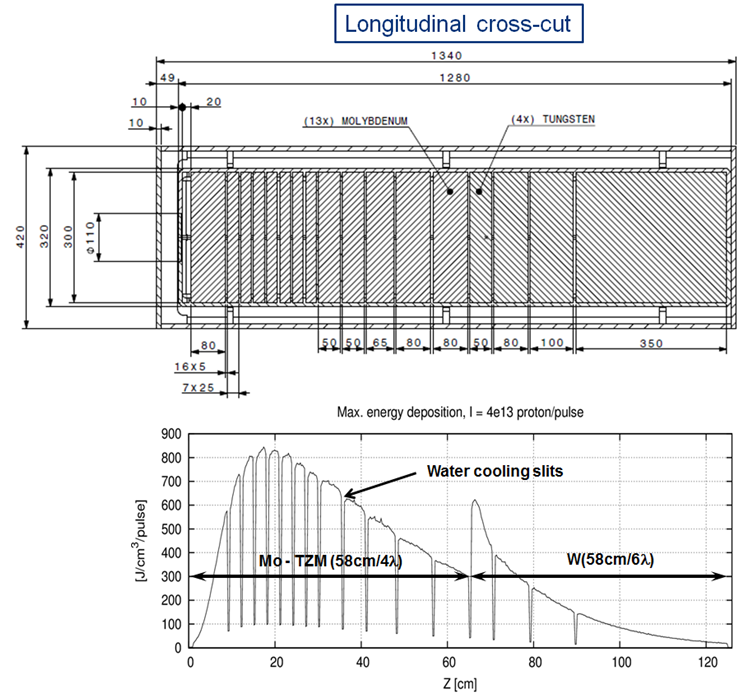}
\caption{(Top) Preliminary design of the proton target configuration. All dimensions are in $\mm$. The right-slanted hatched region in the top drawing shows the TZM blocks and the left-slanted hatched region the tungsten blocks. (Bottom) Peak energy 
deposition in the proton target during a spill of $4\cdot 10^{13}$ protons.}
\label{fig:facility_target_config}
\end{figure}

The very high instantaneous beam power of 
$\sim 2.56\unit{MW}$ per spill of $1.2\sec$ and the average deposited power of $\sim 355\unit{kW}$ over consecutive spills spaced by the SPS cycle of $7.2\sec$ make the design of the 
proton target, its radiological protection, and its cooling very challenging aspects of the facility. Studies show that the required performance may be achieved with a longitudinally 
segmented hybrid target consisting of blocks of four nuclear interaction lengths ($58\cm$) of titanium-zirconium doped molybdenum alloy (TZM, density $10.22\unit{g/cm^3}$ as compared to $10.28\unit{g/cm^3}$ for pure Mo) in the core of the proton shower, 
followed by six nuclear interaction lengths ($58\cm$) of pure tungsten (density $19.3\unit{g/cm^3})$. 
A medium-density material is required in the first half of the target in order to reduce the energy density and create acceptable stresses in the blocks. The blocks are all interleaved with 
$5\mm$ wide slots for water cooling. Tantalum alloy cladding of the TZM and the tungsten blocks is considered in order to prevent corrosion and erosion by the high flow rate of the water cooling. In order to respect the material limits derived from thermo-mechanical stresses, the thickness of each block together with the location of 
each cooling slot has been optimised to provide a relatively uniform energy deposition and sufficient energy extraction. Using FLUKA Monte Carlo simulations~\cite{ref:fluka} and ANSYS finite element analyses, the preliminary target design has been shown to limit the peak power density in 
the target blocks to below $850\unit{J/cm^3}$/spill and compressive stresses below $300\unit{MPa}$ in 
the core of the shower for a $6\mm$ RMS spot size and $30\mm$ single-turn sweep radius. 
Figure~\ref{fig:facility_target_config}~(top) shows the preliminary proton target as designed for the SHiP Technical Proposal~\cite{ref:SHiP_TP}. The total dimensions of the target are $1.2\m$ in length with transverse dimensions of $30~\times~30\unit{cm^2}$. Figure~\ref{fig:facility_target_config}~(bottom) 
shows the maximum energy density per spill of $4\cdot 10^{13}$ protons on target. 

Over the long term, the very high proton cumulated dose alters the physical and mechanical properties 
of the target material such as thermal conductivity and yield strength. First estimates of the radiation damage in terms of the displacement 
per atom, as well as the internal production of hydrogen and helium gas, indicate that the current target design ensures the longevity of the target, but the limited availability 
of data in literature call for accelerated aging studies of the materials with irradiation. A replica target is being designed and built for testing with beam in 2018.

The proton target blocks are assembled in a double-walled vessel.
The inner vessel enforces the high-flow water circulation between the proton target blocks and ensures a 
pressurised water cooling of $15-20\unit{bar}$ in order to avoid water boiling in contact with the target 
blocks. A flow rate of $\sim 180\unit{m^3/h}$ is envisaged. The outer vessel acts as a safety 
hull to contain hypothetical leaks, and is filled with an inert gas to prevent corrosion.


\begin{figure}[tbh]
\centering 
\includegraphics[width=0.8\linewidth]{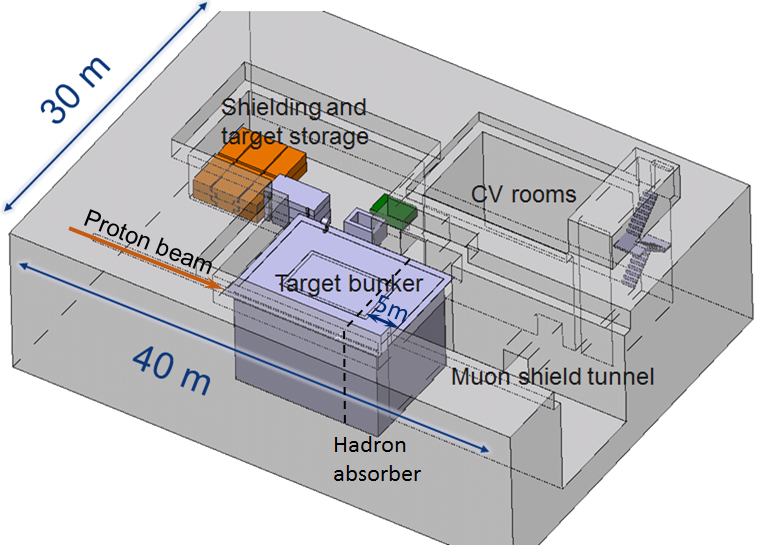}
\caption{Overview of the main components of the target complex. The proton beam line arrives from the left of the target bunker. The target is located in the centre of the target bunker and the first section of the muon shield in terms of the magnetised hadron absorber is integrated in the downstream end of the bunker.}
\label{fig:facility_target_complex}
\end{figure}

\subsection{Preliminary design of the target complex}
\label{sec:target_complex}

An overview of the target complex is shown in Figure~\ref{fig:facility_target_complex}. 
In order to contain the radiation generated by the beam impacting on the proton target, the target is embedded in a $\sim 450\unit{m^3}$ cast-iron bunker. 
The inner part of the cast iron shielding ($\sim 20\unit{m^3}$) is water cooled by means of embedded stainless steel pipes in order to extract the average power of $20\unit{kW}$ which is leaking out of the target during operation. 
The outer part of the shielding is fully passive. The assembly has been designed with emphasis on reliability, remote handling and with the aim of being multi-purpose, 
i.e. allowing exchange of the proton target and the shielding configuration for alternative uses in future experiments. 
To minimise the irradiation of the primary beam line, the upstream shielding has only a limited passage of about $20\cm$ in diameter for the beam vacuum chamber. 
The $5\m$ thick downstream shielding acts as a hadron absorber with the double objective of absorbing the secondary hadrons and the residual non-interacting protons emerging from the target, and significantly reducing
the exposure of the downstream active muon shield to radiation. 
The overall shielding is designed to respect the limits from radiological and environmental protection applicable at CERN. 

A helium-vessel containing high-purity helium gas (>99\%) at atmospheric pressure 
encloses the SHiP proton target and the entire iron shielding. This is required to 
protect the equipment from radiation-accelerated corrosion as well as to avoid the 
production of high-mass radioactive isotopes from secondary neutrons interacting with air. 

\section{Suppression of beam-induced background}

\subsection{Active muon shield}
\label{sec:muon_shield}

The total flux of muons emerging from the proton target with a momentum larger than $1\gevc$ amounts to ${\cal O}(10^{11})$ muons  per spill of $4\cdot 10^{13}$ protons. In order to control the background from random combinations of muons 
producing fake decay vertices in the detector decay volume and from muon deep inelastic scattering 
producing long-lived neutral particles in the surrounding material, and to respect the occupancy limits of 
the sub-detectors, the muon flux in the detector acceptance must be reduced by several orders of magnitude 
over the shortest possible distance. To this end, a muon shield entirely based on magnetic deflection has been 
developed~\cite{ref:muon_shield_engineering, ref:muon_shield_design} (Figure~\ref{fig:SHiP_facility_overview}).

\begin{figure}[t]
\centering
\includegraphics[width=0.6\linewidth]{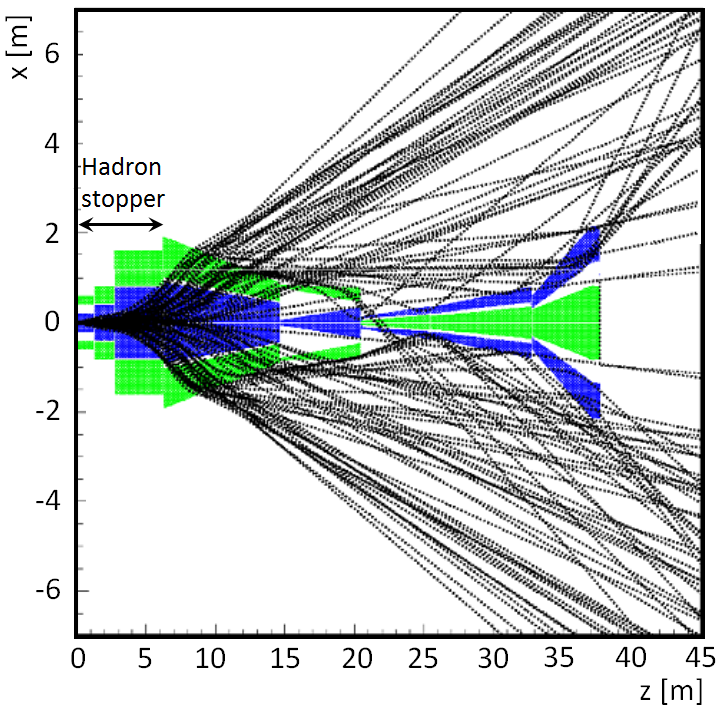}
\caption{Horizontal cross-section of the muon shield magnet configuration at the level of the beam-axis 
(\cite{ref:muon_shield_design}, reproduced here for completeness). The direction up/down of the vertical magnetic 
field is illustrated by the blue/green colour of the iron poles of the magnets. Typical trajectories of 
muons across the momentum spectrum are overlaid.  }
\label{fig:muon_shield}
\end{figure}

Figure~\ref{fig:muon_shield} shows schematically the field configuration of the muon shield magnets.
The first section of the muon shield starts within the hadron absorber with the integration of a 
coil which magnetises the iron shielding block, and continues with a set of freestanding magnets over a
length of $\sim 20\m$. The purpose of the first section is to deflect the positively and negatively charged muons on either side of the beam axis. As shown by the trajectories of the muons in Figure~\ref{fig:muon_shield}, lower momentum muons and muons with larger transverse momenta are swept out of the core field before 
the end of the first section. Due to the return fields, a large fraction of these muons are bent back towards the detector acceptance. 
For this reason, the second section serves two purposes. In addition to providing further bending power to deflect out of acceptance the higher momentum muons, it should also give the lower momentum muons another magnetic kick outwards. This $20\m$ section therefore consists of a series of magnets with the return field close to the z-axis. The residual muons entering the decay volume after the muon shield are mainly due to stochastic processes involving large energy losses and large angle scattering in the muon shield material.

In order to achieve a high magnetic flux of $1.7-1.8\unit{T}$ in the core at low current and with coils of 
small cross-sections, grain-oriented steel is considered as the yoke material for the freestanding 
magnets~\cite{ref:muon_shield_engineering}. The actual field configuration for the entire muon shield 
has been optimised with the help of machine 
learning techniques using a large sample of muons from 
a full GEANT4~\cite{ref:geant4} simulation of $2\cdot 10^{10}$ protons on the SHiP proton target. Engineering studies are underway to 
study the optimal assembly techniques. The total mass of the muon shield magnets is of the order of
$1500\unit{tonnes}$. The current design allows reducing the rate of residual muons above $1\gevc$ reconstructed 
in the SHiP spectrometer to an acceptable rate of ${\cal O}(10^{5})$ per spill.

\subsection{Vacuum vessel}
\label{sec:vac_vessel}

\begin{figure}[t]
\centering
\includegraphics[width=0.9\linewidth]{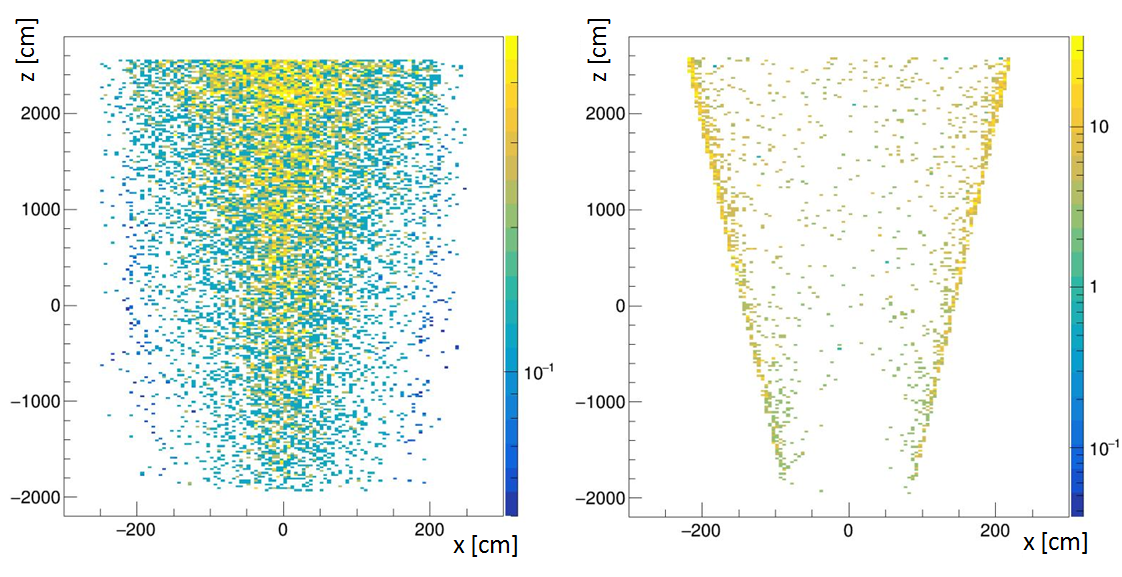}
\caption{Vertex distribution of signal 
candidates produced by neutrino interactions from $2\cdot 10^{20}$ protons on target assuming air at atmospheric pressure in the fiducial volume with
a soft selection for heavy neutral leptons (left), as compared to the situation with a vacuum vessel evacuated down to a pressure of $10^{-3}\unit{bar}$ (right).}
\label{fig:neutrino_vertices}
\end{figure}

Deep inelastic neutrino-nucleon scattering in the detector volume leads to 
background events through the production of $V^0$ particles ($K_L, K_S, \Lambda$) whose decay
mimic the topology and modes of the hidden particle decays.
With $2\cdot 10^{20}$ protons on target, a flux of $\sim 4.5\cdot 10^{18}$ neutrinos and $\sim 3\cdot 10^{18}$ 
anti-neutrinos are expected within the angular acceptance of the SHiP detector. The flux is dominated by muon 
neutrinos coming from the decays of pions and kaons produced in the proton target. Neutrinos 
from decays of charm and beauty hadrons constitute 
$\sim 10\%$ of the total neutrino flux. Figure~\ref{fig:neutrino_vertices}~(left) shows the vertex distribution of signal 
candidates produced by neutrino interactions assuming air at atmospheric pressure in the fiducial volume, and no surrounding vessel structure. A soft selection for heavy neutral leptons based on finding a vertex in the fiducial volume and no activity in the upstream detectors is applied. In these conditions, a total number of $2.5\cdot 10^{3}$ candidate events are expected within
the acceptance for $2\cdot 10^{20}$ protons on target. The events are largely concentrated along the centre with 
small reconstructed impact parameters at the proton target. To achieve the required level of neutrino background rejection, the 
fiducial volume is therefore contained in a
vacuum vessel (Figure~\ref{fig:SHiP_facility_overview}) which is evacuated down to a pressure 
of ${\cal{O}}(10^{-3})\unit{bar}$. Figure~\ref{fig:neutrino_vertices}~(right) shows the vertex distribution of signal candidates at this pressure.
In these conditions, $1.4\cdot10^{4}$ candidate events are expected within the fiducial volume with the same soft
selection for $2\cdot 10^{20}$ protons on target, mainly produced through neutrino interactions with the vessel walls.
Even if the total number of neutrino interactions are larger due to the vessel material, almost all candidate events are in this case easily rejected by using criteria based on the reconstructed impact parameter at the 
proton target. In addition, residual neutrino interactions as well as muon deep inelastic interactions with the vessel structure
are further suppressed by instrumenting the entire decay volume walls with a background tagger system and detecting the 
additional activity associated with the interactions. Simulation studies show that no background events remain after applying
these criteria~\cite{ref:SHiP_TP}.

\begin{figure}[tbh]
\centering
\includegraphics[width=0.38\linewidth]{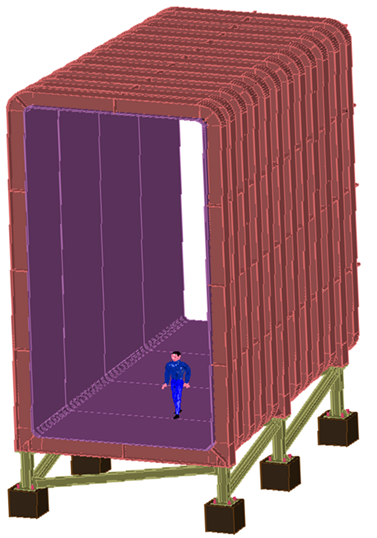}
\caption{Cross-sectional view of the vacuum vessel which provides a pressure of ${\cal O}(10^{-3})\unit{bar}$ in the decay volume. The design has been optimised in order for the wall to be as light and as slim as possible, and to incorporate a detector system which tags background events.}  
\label{fig:vacuum_vessel}
\end{figure}

The SHiP decay vessel consists of the $\sim 50\m$ decay volume constructed in S355JO(J2/K2)W Corten steel with upstream 
outer dimensions of $2.4\times 4.5\ma$ and downstream outer dimensions of $5\times 10\ma$. The design of the vessel 
wall is based on an optimisation aiming at producing a 
structure as light as possible and as slim as 
possible in order to stay within the boundaries of the deflected muon flux whilst maintaining the required acceptance.
At the same time, the optimisation also accounts for the structural safety norms allowing access to the underground hall while under vacuum 
and the earthquake loads in the region. 
Figure~\ref{fig:vacuum_vessel} shows the structure of the decay volume. The preliminary design consists of a $30\mm$ thick
continuous inner steel sheet acting as vacuum liner, supported azimuthally by welded T-shaped beams with a steel thickness of $15\mm$ and a 
height varying from $300\mm$ to $450\mm$. The structure is further reinforced by longitudinal stiffening profiles
between the azimuthal beams. 

Two options are considered for the surrounding background tagger, either a liquid or a plastic scintillator. 
For the liquid option, the scintillator is integrated within the decay volume structure with an extra $8\mm$ steel sheet welded to the 
azimuthal beams and stiffening profiles, while for the plastic scintillator option, it is attached directly to the structure. 
The decay volume is directly connected to the $\sim 10\m$ downstream spectrometer vacuum section, which is made 
out of austenitic steel since it runs through the spectrometer magnet and houses the four tracker stations of 
straw tubes built using the same technology from the NA62 experiment~\cite{ref:NA62}. The preliminary design considers extruded aluminum profiles with a material budget 
equivalent to $0.8$ radiation lengths for the upstream and downstream windows.

\section{Conclusions}
\label{sec:conclusions}

The SHiP experimental facility will provide a unique experimental platform for physics at the intensity
frontier which is complementary to both the searches for new physics at the energy frontier and the direct searches for cosmic Dark Matter.  CERN's accelerator complex makes for an ideal siting for the experimental facility. The assumed availability of $2\cdot 10^{20}$ protons on target at $400\gevc$ in about five years of nominal operation and an environment of extremely low background compares favourably with the potential of other existing facilities.

The two-fold SHiP apparatus is sensitive both to decays and to scattering signatures, and is able to probe a wide variety of models with light long-lived exotic particles in a largely unexplored domain of very weak couplings and masses up to ${\cal O}(10)\gevcc$. This puts it in a unique position worldwide to resolve several of the major observational puzzles of particle physics and cosmology. In addition, the same facility enables the study of interactions of tau neutrino and anti-tau neutrinos, as well as neutrino-induced charm production by all neutrino species.  A more recent investigation also shows that an additional detector on the SHiP beam-line with a proton target consisting of thin wires and operating in parallel would allow a search for lepton flavour violating tau lepton decays at a sensitivity that could be highly competitive with projections of approved experiments.

The experimental facility presents a number of technological challenges to the beam delivery, the proton target system, and the reduction of beam-induced background.  As reported, in-depth studies and prototyping are already well underway for all of the critical components.  Taking into account the required R\&D and construction, and the accelerator schedule at CERN, we plan to commission and perform the pilot run for the SHiP experiment when the SPS resumes operation after LHC's third long shutdown for maintenance and upgrades.  

\section{Acknowledgements}
The SHiP Collaboration wishes to thank the Castaldo company (Naples, Italy) for their contribution to the development studies of the decay vessel. 
The support from the National Research Foundation of Korea with grant numbers of 2018R1A2B2007757, 2018R1D1A3B07050649, 2018R1D1A1B07050701, 2017R1D1A1B03036042, 2017R1A6A3A01075752, 2016R1A2B4012302, and 2016R1A6A3A11930680 is acknowledged.
The support from the Russian Foundation for Basic Research, grant 17-02-00607, and the support from the TAEK of Turkey
are acknowledged.

\end{document}

%% file: lhcb-symbols-def.tex
\usepackage{xspace} 
\usepackage{upgreek}

\def\MagUp {\mbox{\em Mag\kern -0.05em Up}\xspace}

\ifthenelse{\boolean{uprightparticles}}%
{

 \def\PDelta      {\ensuremath{\Delta}\xspace}                 
 \def\PXi      {\ensuremath{\Xi}\xspace}                 
 \def\PLambda      {\ensuremath{\Lambda}\xspace}                 
 \def\PSigma      {\ensuremath{\Sigma}\xspace}                 
 \def\POmega      {\ensuremath{\Omega}\xspace}                 
 \def\PUpsilon      {\ensuremath{\Upsilon}\xspace}                 
 
 \def\PB      {\ensuremath{\mathrm{B}}\xspace}                 
                  
 \def\PD      {\ensuremath{\mathrm{D}}\xspace}

 \def\PK      {\ensuremath{\mathrm{K}}\xspace}

 \def\Pi      {\ensuremath{\mathrm{i}}\xspace}

}
{

 \mathchardef\PDelta="7101
 \mathchardef\PXi="7104
 \mathchardef\PLambda="7103
 \mathchardef\PSigma="7106
 \mathchardef\POmega="710A
 \mathchardef\PUpsilon="7107
                  
 \def\PB      {\ensuremath{B}\xspace}                 
                  
 \def\PD      {\ensuremath{D}\xspace}

 \def\PK      {\ensuremath{K}\xspace}

 \def\Pi      {\ensuremath{i}\xspace}

}

\makeatletter
\ifcase \@ptsize \relax
  \newcommand{\miniscule}{\@setfontsize\miniscule{4}{5}}
\or
  \newcommand{\miniscule}{\@setfontsize\miniscule{5}{6}}
\or
  \newcommand{\miniscule}{\@setfontsize\miniscule{5}{6}}
\fi
\makeatother

\DeclareRobustCommand{\optbar}[1]{\shortstack{{\miniscule (\rule[.5ex]{1.25em}{.18mm})}
  \\ [-.7ex] $#1$}}

  \def\Kbar    {{\kern 0.2em\overline{\kern -0.2em \PK}{}}\xspace}

\def\KorKbar    {\kern 0.18em\optbar{\kern -0.18em K}{}\xspace}

  \def\Dbar    {{\kern 0.2em\overline{\kern -0.2em \PD}{}}\xspace}

\def\DorDbar    {\kern 0.18em\optbar{\kern -0.18em D}{}\xspace}

\def\Bbar    {{\ensuremath{\kern 0.18em\overline{\kern -0.18em \PB}{}}}\xspace}

\def\BorBbar    {\kern 0.18em\optbar{\kern -0.18em B}{}\xspace}

  \def\Y#1S{\ensuremath{\PUpsilon{(#1S)}}\xspace}

\def\Lbar        {{\ensuremath{\kern 0.1em\overline{\kern -0.1em\PLambda}}}\xspace}
\def\LorLbar    {\kern 0.18em\optbar{\kern -0.18em \PLambda}{}\xspace}

\def\AT#1     {\ensuremath{A_{\mathrm{T}}^{#1}}\xspace}           

\def\C#1      {\ensuremath{\mathcal{C}_{#1}}\xspace}                       
\def\Cp#1     {\ensuremath{\mathcal{C}_{#1}^{'}}\xspace}                    
\def\Ceff#1   {\ensuremath{\mathcal{C}_{#1}^{\mathrm{(eff)}}}\xspace}        
\def\Cpeff#1  {\ensuremath{\mathcal{C}_{#1}^{'\mathrm{(eff)}}}\xspace}       
\def\Ope#1    {\ensuremath{\mathcal{O}_{#1}}\xspace}                       
\def\Opep#1   {\ensuremath{\mathcal{O}_{#1}^{'}}\xspace}                    

\newcommand{\unit}[1]{\ensuremath{\mathrm{ \,#1}}\xspace}          

\newcommand{\tev}{\ifthenelse{\boolean{inbibliography}}{\ensuremath{~T\kern -0.05em eV}\xspace}{\ensuremath{\mathrm{\,Te\kern -0.1em V}}}\xspace}
\newcommand{\gev}{\ensuremath{\mathrm{\,Ge\kern -0.1em V}}\xspace}
\newcommand{\mev}{\ensuremath{\mathrm{\,Me\kern -0.1em V}}\xspace}
\newcommand{\kev}{\ensuremath{\mathrm{\,ke\kern -0.1em V}}\xspace}
\newcommand{\ev}{\ensuremath{\mathrm{\,e\kern -0.1em V}}\xspace}
\newcommand{\gevc}{\ensuremath{{\mathrm{\,Ge\kern -0.1em V\!/}c}}\xspace}
\newcommand{\mevc}{\ensuremath{{\mathrm{\,Me\kern -0.1em V\!/}c}}\xspace}
\newcommand{\gevcc}{\ensuremath{{\mathrm{\,Ge\kern -0.1em V\!/}c^2}}\xspace}
\newcommand{\gevgevcccc}{\ensuremath{{\mathrm{\,Ge\kern -0.1em V^2\!/}c^4}}\xspace}
\newcommand{\mevcc}{\ensuremath{{\mathrm{\,Me\kern -0.1em V\!/}c^2}}\xspace}

\def\m    {\ensuremath{\mathrm{ \,m}}\xspace}
\def\ma   {\ensuremath{{\mathrm{ \,m}}^2}\xspace}
\def\cm   {\ensuremath{\mathrm{ \,cm}}\xspace}

\def\mm   {\ensuremath{\mathrm{ \,mm}}\xspace}

\def\sec  {\ensuremath{\mathrm{{\,s}}}\xspace}

\def\ps   {\ensuremath{{\mathrm{ \,ps}}}\xspace}

\def\gsim{{~\raise.15em\hbox{$>$}\kern-.85em
          \lower.35em\hbox{$\sim$}~}\xspace}
\def\lsim{{~\raise.15em\hbox{$<$}\kern-.85em
          \lower.35em\hbox{$\sim$}~}\xspace}

\def\tell1  {TELL1\xspace}
\def\ukl1   {UKL1\xspace}

